\documentclass[aps,preprint,nofootinbib]{revtex4}

\usepackage{graphicx}

\begin{document}

\title{$B\to K$ Transition Form Factor with Tensor Current within the $k_T$ Factorization Approach}
\author{Dai-Min Zeng, Xing-Gang Wu\footnote{email: wuxg@cqu.edu.cn} and Zhen-Yun Fang}
\address{Department of Physics, Chongqing University, Chongqing 400044,
P.R. China}

\begin{abstract}
In the paper, we apply the $k_T$ factorization approach to deal with
the $B\to K$ transition form factor with tensor current in the large
recoil regions. Main uncertainties for the estimation are discussed
and we obtain $F_T^{B\to K}(0)=0.25\pm0.01\pm0.02$, where the first
error is caused by the uncertainties from the pionic wave functions
and the second is from that of the B-meson wave functions. This
result is consistent with the light-cone sum rule results obtained
in the literature. \\

\noindent {\bf PACS numbers:} 12.38.Aw, 12.38.Lg, 13.20.He
\end{abstract}

\maketitle

There is an increasing demand for more reliable QCD calculations of
the heavy-to-light form factors, which plays a complementary role in
determination of the fundamental parameters of the standard model
and in developing the QCD theory. The $B\to K$ transition form
factors $F_+^{B\to K}(q^2)$ and $F_0^{B\to K}(q^2)$ have been
studied up to ${\cal O}(1/m^2_b)$ in the large recoil region within
the $k_T$ factorization approach \cite{wu}, where the B-meson wave
functions $\Psi_B$ and $\bar\Psi_B$ that include the three-Fock
states' contributions are adopted and the transverse momentum
dependence for both the hard scattering part and the
non-perturbative wave function, the Sudakov effects and the
threshold effects are included to regulate the endpoint singularity
and to derive a more reliable PQCD result. The rich flavor changing
neutral current process $B\to K l^+ l^-$ have attracted people's
attentions recently, since this decay provides potential testing
grounds for the standard model at loop level and and is a hopeful
channel to probe the new physics beyond the standard model, c.f.
Ref.\cite{bsm} and references therein. A better understanding of the
rare semi-leptonic decay $B\to K l^+ l^-$ \cite{belle} needs a
better understanding of its key component, i.e. the $B \to K$
transition form factor with tensor current $F_T^{B\to K}(q^2)$. So,
in addition to $F_+^{B\to K}(q^2)$ and $F_0^{B\to K}(q^2)$, it is
also very interesting to study the properties of $F_T^{B\to
K}(q^2)$, which is the purpose of the present letter.

\begin{figure}
\centering
\includegraphics[width=0.6\textwidth]{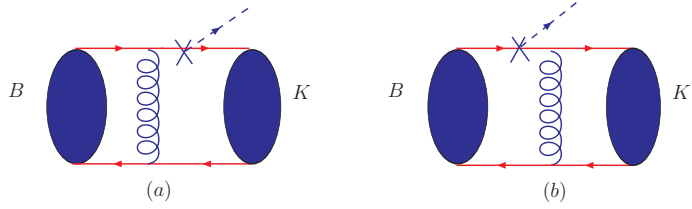}
\caption{Lowest order hard-scattering kernel for the $B\to K$
transition form factor, where the cross denotes an appropriate gamma
matrix ($\sigma_{\mu\nu}$).} \label{figbk}
\end{figure}

The $B\to K$ transition form factor $F_T^{B\to K}(q^2)$ is defined
as follows:
\begin{equation}\label{eq:bk2}
\langle K(P_K)|\bar s \sigma_{\mu\nu}q^{\nu}b|\bar
B(P_B)\rangle=i\frac{F_T^{B\to K}(q^2)}{M_B
+M_K}\left[q^2(P_B+P_K)_{\mu}-(M_B^2-M_K^2)q_{\mu}\right] ,
\end{equation}
where the momentum transfer $q=P_{B} - P_{K}$. The amplitude for the
$B\to K$ transition form factor can be factorized into the
convolution of the wave functions for the respective hadrons with
the hard-scattering amplitude. In the large recoil regions, the
$B\to K$ transition form factor is dominated by a single gluon
exchange in the lowest order, whose Feynman diagram is shown in
Fig.(\ref{figbk}). In the hard scattering kernel, the transverse
momentum in the denominators are retained to regulate the endpoint
singularity. The masses of the light quarks are neglected. The terms
proportional to $\mathbf{k}_\bot^2$ or $\mathbf{l}_\bot^2$ in the
numerator are dropped, which are power suppressed compared to other
${\cal O}(M_B^2)$ terms. Under these treatment, the Sudakov form
factor from $k_T$ resummation can be introduced into the PQCD
factorization theorem without breaking the gauge invariance
\cite{li1}. As for the $B\to K$ transition form factor $F_T^{B\to
K}(q^2)$, it can be written in the transverse configuration
$b$-space by properly including the Sudakov form factors and the
threshold resummation effects:
\begin{eqnarray}
F_T^{B\to K}(q^2) &=& \frac{\pi C_F}{N_c} f_{K}f_B M_B^2\int d\xi
dx\int b_B db_B~ b_K db_K~ \alpha_s(t)
\times\exp[-S(x,\xi,b_K,b_B;t)] \nonumber\\
&&\times S_t(x)S_t(\xi)\Bigg \{ \Bigg [ \Psi_K(x, b_K)\left (
\Psi_B(\xi, b_B)-\bar\Psi_B(\xi, b_B) \right )+
\frac{m_0^p}{M_B}\Psi_p(x, b_K )\cdot\nonumber \\
&&\left(\frac{1}{\eta}\bar\Psi_B(\xi,b_B) -x\Psi_B(\xi,b_B)\right)
+\frac{m_0^p}{M_B}\frac{\Psi'_\sigma(x,b_K )}{6}
\left(\frac{x\eta+2}{\eta}\Psi_B(\xi,b_B)\right.\nonumber\\
&&\left. -\frac{1}{\eta}\bar\Psi_B(\xi,b_B)
\right)+\frac{m_0^p}{M_B} \frac{\Psi_\sigma(x,b_K
)}{6}\Psi_B(\xi,b_B)\Bigg ] h_1(x,\xi,b_K,b_B)- \frac{m_0^p}{M_B}
\frac{\Psi_\sigma(x,b_K )}{6}\nonumber \\
&& [M_B\Delta(\xi,b_B)]h_2(x,\xi,b_K,b_B) +\Bigg [ \Psi_K(x,
b_K )\left (\frac{\Delta(\xi,b_B)}{M_B}-\xi\Psi_B(\xi,b_B)\right )+\nonumber\\
&& 2\frac{m_0^p}{M_B}\Psi_p(x,b_K )\left(
\Psi_B(\xi,b_B)-\frac{\xi}{\eta}\bar\Psi_B(\xi,b_B)\right ) \Bigg ]
h_1(\xi,x,b_B,b_K ) \Bigg \}, \label{fbcT}
\end{eqnarray}
where
\begin{eqnarray}
h_1(x,\xi,b_K ,b_B)&=&K_0(\sqrt{\xi x\eta}~M_B b_B)\Bigg [
\theta(b_B-b_K )I_0(\sqrt{x\eta}~M_Bb_K )K_0(\sqrt{x\eta}~M_B b_B) \nonumber \\
&&+\theta(b_K -b_B)I_0(\sqrt{x\eta}~M_Bb_B) K_0(\sqrt{x\eta}~M_B b_K ) \Bigg ], \\
h_2(x,\xi,b_K ,b_B)&=&\frac{b_B}{2\sqrt{\xi xy}M_B}K_{1}(\sqrt{\xi
x\eta}~M_B b_B)\Bigg [ \theta(b_B-b_K )I_0(\sqrt{x\eta}~M_Bb_K ) K_0(\sqrt{x\eta}~M_B b_B) \nonumber \\
&&+\theta(b_K -b_B)I_0(\sqrt{x\eta}~M_Bb_B) K_0(\sqrt{x\eta}~M_B b_K
) \Bigg ].
\end{eqnarray}
The functions $I_i$ ($K_i$) are the modified Bessel functions of the
first (second) kind with the $i$-{\it th} order. The angular
integrations in the transverse plane have been performed. The factor
$\exp(-S(x,\xi,b_K ,b_B;t))$ contains the Sudakov logarithmic
corrections and the renormalization group evolution effects of both
the wave functions and the hard scattering amplitude,
\begin{equation}
S(x,\xi,b_K,b_B;t)= \left[s(x,b_K ,M_b)+s(\bar{x},b_K
,M_b)+s(\xi,b_B,M_b)
-\frac{1}{\beta_{1}}\ln\frac{\hat{t}}{\hat{b}_K}
-\frac{1}{\beta_{1}}\ln\frac{\hat{t}}{\hat{b}_B} \right],
\end{equation}
where ${\hat t}={\rm ln}(t/\Lambda_{QCD})$, ${\hat b}_B ={\rm
ln}(1/b_B\Lambda_{QCD})$, ${\hat b}_K ={\rm ln}(1/b_K \Lambda_{QCD})
$ and $s(x,b,Q)$ is the Sudakov exponent factor, whose explicit form
up to next-to-leading log approximation can be found in
Ref.\cite{liyu}. $S_t(x)$ and $S_t(\xi)$ come from the threshold
resummation effects and here we take a simple parametrization
proposed in Refs.\cite{li1,kls},
\begin{equation}
S_t(x)=\frac{2^{1+2c}\Gamma(3/2+c)}{\sqrt{\pi}\Gamma(1+c)}
[x(1-x)]^c\;,
\end{equation}
where the parameter $c$ is determined around $0.3$ for the present
case. The hard scale $t$ in $\alpha_s(t)$ and the Sudakov form
factor might be varied for the different hard scattering parts and
here we need two $t_i$\cite{li1,lucai}, whose values are chose as
the largest scale of the virtualitiies of internal particles, i.e.
\begin{equation}
t_1={\rm MAX}\left(\sqrt{x\eta}M_B,1/b_K ,1/b_B\right),\; t_2={\rm
MAX} \left(\sqrt{\xi\eta}M_B,1/b_K ,1/b_B\right).
\end{equation}
The Fourier transformation for the transverse part of the wave
function is defined as
\begin{equation}\label{fourier}
\Psi(x,\mathbf{b})=\int_{|\mathbf{\mathbf{k}}|<1/b}
d^2\mathbf{k}_\perp\exp\left(-i\mathbf{k}_\perp
\cdot\mathbf{b}\right)\Psi(x,\mathbf{k}_\perp),
\end{equation}
where $\Psi$ stands for $\Psi_K$, $\Psi_p$, $\Psi_\sigma$, $\Psi_B$,
$\bar\Psi_B$ and $\Delta$, respectively. The upper edge of the
integration $|\mathbf{k}_\perp|<1/b$ is necessary to ensure that the
wave function is soft enough \cite{huang2}.

In the numerical calculations, we use
\begin{eqnarray}
\Lambda^{(n_f=4)}_{\over{MS}}=250MeV,\;\;  f_B=190MeV,\;\; M_B=5.279
GeV .
\end{eqnarray}
Further more, we need to know the non-perturbative wave functions
for the B meson and kaon. Here we take the models as adopted in
Ref.\cite{wu} to do our calculation, where only the kaon twist-2
wave function should be slightly changed to include the second
Gegenbauer moment $a^K_2$'s effect as suggested in
Ref.\cite{wuhuang}, i.e.
\begin{equation}\label{model}
\Psi_{K}(x,\mathbf{k}_\perp) = [1+B_K C^{3/2}_1(2x-1)+C_K
C^{3/2}_2(2x-1)]\frac{A_K}{x(1-x)} \exp \left[-\beta_K^2
\left(\frac{k_\perp^2+m_q^2}{x}+ \frac{k_\perp^2+m_s^2}
{1-x}\right)\right],
\end{equation}
where $q=u,\; d$, $C^{3/2}_1(1-2x)$ is the Gegenbauer polynomial.
The constitute quark masses are set to be: $m_q=0.30{\rm GeV}$ and
$m_s=0.45{\rm GeV}$. It can be found that the $SU_f(3)$ symmetry is
broken by a non-zero $B_K$ and by the mass difference between the
$s$ quark and $u$ (or $d$) quark in the exponential factor. So the
$SU_f(3)$ broken effects to the form factor are naturally included
into our discussions. We will take $a_{1}^{K}(1GeV)=0.05\pm0.02$ to
determine the wave function $\Psi_K$. As for $a^K_2$, since it is
still determined with large uncertainty and for convenience, we fix
its value to be $a^K_2(1GeV)=0.115$ \cite{sumrule}. The four
parameters $A_K$, $B_K$, $C_K$ and $\beta_K$ can be determined by
its first two Gegenbauer moments $a^K_1$ and $a^K_2$, the constraint
$\langle \mathbf{k}_\perp^2 \rangle^{1/2}_K \approx 0.350{\rm GeV}$
\cite{gh} and the normalization condition $\int^1_0 dx
\int_{k_\perp^2<\mu_0^2} \frac{d^{2}{\bf
k}_{\perp}}{16\pi^3}\Psi_K(x,{\bf k}_{\perp}) =1$. For example, we
have $A_K(\mu_b)=252.044GeV^{-1}$, $B_K(\mu_b)=0.09205$,
$C_K(\mu_b)=0.05250$ and $\beta_K=0.8657GeV^{-1}$ for the case of
$a^K_1(1GeV)=0.05$ and $a^K_2(1GeV)=0.115$. As for the $B$ meson
wave functions, we adopt the simple model raised in Ref.\cite{hqw}
to do our discussion:
\begin{equation}\label{newmodel1}
\Psi^{+}_{B}(\xi,b_B)=(16\pi^3)\frac{M_B^2\xi}{\omega_0^2}\exp
\left( -\frac{M_B\xi}{\omega_0}\right) \Big(\Gamma[\delta]
J_{\delta-1} [\kappa] +(1-\delta)\Gamma[2-\delta]
J_{1-\delta}[\kappa]\Big)\left( \frac{\kappa}{2} \right)^{1-\delta}
\end{equation}
and
\begin{equation}
\Psi^{-}_{B}(\xi,b_B)=(16\pi^3)\frac{M_B}{\omega_0}\exp \left(
-\frac{M_B \xi}{\omega_0}\right)\Big(\Gamma[\delta] J_{\delta-1}
[\kappa] +(1-\delta)\Gamma[2-\delta] J_{1-\delta}[\kappa]\Big)\left(
\frac{\kappa}{2} \right)^{1-\delta} ,\label{newmodel2}
\end{equation}
which satisfy the normalization $\int \frac{d\xi d^{2}{\bf
k}_{\perp}}{16\pi^3} \Psi^{\pm}_{B}(\xi,\mathbf{k}_\perp) =1$.
$\omega_0=2\bar\Lambda/3$, $\bar\xi=\bar\Lambda/M_B$,
$\kappa=\theta(2\bar\xi-\xi) \sqrt{\xi (2\bar\xi-\xi)} M_B b_B $.
According to the definitions, we have
$\Psi_{B}(\xi,b_B)=\Psi^{+}_{B}(\xi,b_B)$,
$\bar\Psi_{B}(\xi,b_B)=\Psi^{+}_{B}(\xi,b_B)-\Psi^{-}_{B}(\xi,b_B)$
and $\Delta(\xi,b_B)=-M_B\int^\xi_0 d\xi' \bar\Psi_B(\xi',b_B)$. The
$B$-meson wave function Fock state expansion depends on two
phenomenological parameters $\bar\Lambda$ and $\delta$. We will take
$\bar\Lambda\in[0.50, 0.55] GeV$ and $\delta\in [0.25,0.30]$ to
study their uncertainties to the form factor $F^{B\to
K}_{T}(q^{2})$, which is determined by comparing the PQCD results of
$B\to\pi$ form factor with the QCD LCSR results and lattice QCD
calculations \cite{hwbpi}.

\begin{table}
\centering
\begin{tabular}{c|c|c|c}
\hline\hline ~~~ ~~~ &  ~~~this work~~~ & ~~~LCSR \cite{alexander} ~~~ & ~~~LCSR \cite{sumrule} \\
\hline
$F^{B\to K}_{T}(0)$ & $0.25\pm0.01\pm0.02$ & $0.27\pm0.04$ &  $0.321\pm0.046$ \\
\hline\hline
\end{tabular}
\caption{Simple comparison of $F^{B\to K}_{T}(q^{2}=0)$ calculated
within the present adopted $k_T$ factorization approach and the LCSR
approach \cite{alexander,sumrule}. }\label{tab}
\end{table}

By varying the undetermined parameters, such as $a^K_1$,
$\bar\Lambda$ and $\delta$, we compare our results of $F^{B\to
K}_{T}(q^{2}=0)$ with those derived from the QCD light-cone sum
rules in TAB.\ref{tab}. We obtain $F^{B\to
K}_{T}(q^{2}=0)=0.25\pm0.01\pm0.02$, where the center value is
obtained by setting $a^K_1(1GeV)=0.05$, $\bar\Lambda=0.525GeV$ and
$\delta=0.275$, and the first error comes from the uncertainty of
$a^K_1$ and the second comes from that of $\bar\Lambda$ and
$\delta$. Our result shows a good agreement with the LCSR result of
Ref.\cite{alexander}, and both of which roughly agree with the
result of Ref.\cite{sumrule} within theoretical errors. Further
more, it can be found that the PQCD results can match with the LCSR
results for small $q^2$ region, e.g. $q^2<10GeV^2$. Then by
combining the PQCD results with the LCSR results, we can obtain a
consistent analysis of the form factor within the large and the
intermediate energy regions \cite{hwbpi,wuhuang}. Inversely, if the
PQCD approach must be consistent with the LCSR approach, then we can
obtain some constraints to the undetermined parameters within both
approaches.

It may be interesting to know how the undermined parameters, such as
$a^K_1$, $\bar\Lambda$ and $\delta$, affect the form factor $F^{B\to
K}_{T}(q^{2})$.

\begin{figure}
\centering
\includegraphics[width=0.45\textwidth]{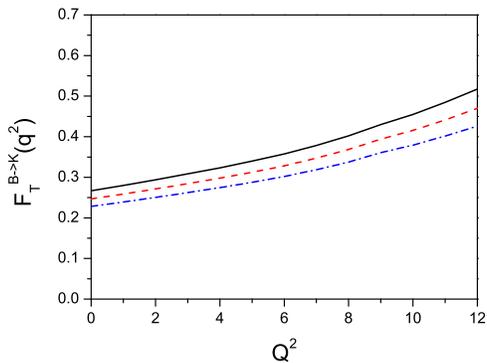}
\caption{PQCD results for the $B\to K$ form factors $F_T^{B\to
K}(q^2)$ with fixed $\delta=0.275$ and $a^{K}_{1}(1GeV)=0.05$. The
solid line, the dashed line and the dash-dot line are for the cases
of $\bar\Lambda=0.500GeV$, $0.525GeV$ and $0.550GeV$ respectively.}
\label{lambda}
\end{figure}

\begin{figure}
\centering
\includegraphics[width=0.45\textwidth]{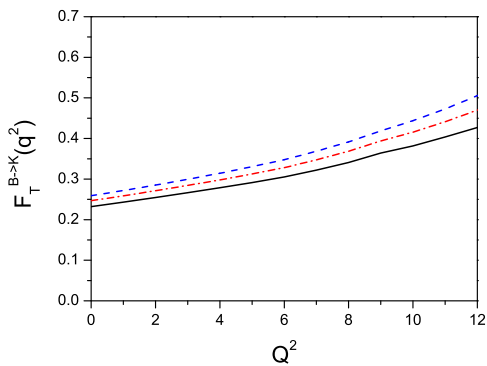}
\caption{PQCD results for the $B\to K$ form factors $F_T^{B\to
K}(q^2)$ with $\bar\Lambda=0.525GeV$ and $a^{K}_{1}(1GeV)=0.05$. The
solid line, the dash-dot line and the dashed line stand for the
cases of $\delta=0.25$, $0.275$ and $0.30$ respectively.}
\label{delta}
\end{figure}

We first discuss the uncertainties of $F^{B\to K}_{T}(q^{2})$ arise
from the B-meson wave function, i.e. to discuss the uncertainties
from $\bar\Lambda$ and $\delta$. For such purpose, we fix the kaonic
wave function by setting $a^{K}_{1}(1GeV)=0.05$. The transition form
factor $F^{B\to K}_{T}(q^{2})$ with fixed $\delta=0.275$ is shown in
Fig.(\ref{lambda}), where $\bar\Lambda$ varies within the region of
$[0.50GeV,0.550GeV]$. Fig.(\ref{lambda}) shows that $F^{B\to
K}_{T}(q^{2})$ decreases with the increment of $\bar\Lambda$. While
the transition from factor $F^{B\to K}_{T}(q^{2})$ with fixed
$\bar\Lambda=0.525GeV$ is shown in Fig.(\ref{delta}), where $\delta$
varies within the region of $[0.25,0.30]$. Fig.(\ref{delta}) shows
that $F^{B\to K}_{T}(q^{2})$ increases with the increment of
$\delta$. As a whole, it can be found that by varying
$\bar\Lambda\in[0.50GeV,0.550GeV]$ and $\delta\in[0.25,0.30]$, then
it will cause about $10\%$ uncertainty to $F^{B\to K}_{T}(q^{2})$.

\begin{figure}
\centering
\includegraphics[width=0.45\textwidth]{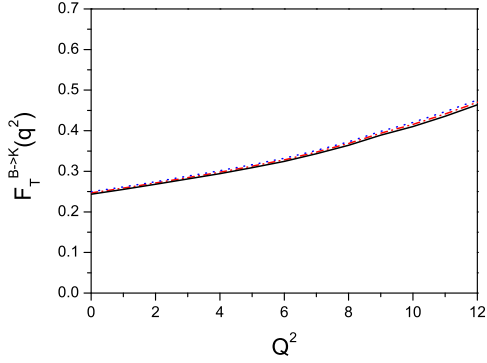}
\caption{PQCD results for the $B\to K$ form factors $F_T^{B\to
K}(q^2)$ with $\bar\Lambda=0.525GeV$ and $\delta=0.275$. The solid
line, the dash-dot line and the dotted line stand for
$a^{K}_{1}(1GeV)=0.03,0.05$ and $0.07$ respectively.} \label{ak1}
\end{figure}

Second, we discuss the properties of $F^{B\to K}_{T}(q^{2})$ caused
by the twist-2 wave function $\Psi_{K}$, ie. by the value of
$a^{K}_{1}(1GeV)$. For this purpose, we fix the B-meson wave
functions by setting $\bar\Lambda=0.525GeV$ and $\delta=0.275$. We
show the $B\to K$ transition form factor $F^{B\to K}_{T}(q^{2})$ in
Fig.(\ref{ak1}) with $a^{K}_{1}(1GeV)=0.03$, $0.05$ and $0.07$
respectively. It is found that $F^{B\to K}_{T}(q^{2})$ will slightly
increase with the increment of $a^{K}_{1}(1GeV)$. And by varying
$a^{K}_{1}(1GeV)\in [0.03, 0.07]$, it will cause about $1\%$
uncertainty to $F^{B\to K}_{T}(q^{2})$.

In summary: we have applied the $k_{T}$ factorization approach to
calculate the $B\to K$ transition form factor $F^{B\to
K}_{T}(q^{2})$ up to order $(1/m^2_b)$, where the transverse
momentum dependence for the wave function, the Sudakov effects and
the threshold effects are included to regulate the endpoint
singularity and to derive a more reasonable result. By varying the
undetermined parameters, such as $a^K_1$, $\bar\Lambda$ and
$\delta$, within the reasonable reasonable regions, we obtain
$F^{B\to K}_{T}(0) = 0.25\pm0.01\pm0.02$. It shows that the $k_T$
factorization can be applied to calculate the form factors in the
large recoil regions. Together with the newly developed PQCD results
of $F^{B\to K}_{+}(q^2)$ and $F^{B\to K}_{0}(q^2)$ \cite{wu}, one
can achieve a full understanding of the these three $B\to K$ form
factors in the large recoil regions. Furthermore, in combination of
the LCSR results, one can know well the $B\to K$ form factor
$F^{B\to K}_{T}(q^2)$ in the large and intermediate energy regions
and then to derive a better understanding of the rare semi-leptonic
decay $B\to K l^+ l^-$ \cite{belle}. \\

\noindent{\bf Acknowledgement}: This work was supported in part by
the Natural Science Foundation of China (NSFC), by the Grant from
Chongqing University and by the National Basic Research Programme of
China under Grant NO. 2003CB716300.

\end{document}